\documentclass[twocolumn]{article}%
\usepackage{amsmath}%
\usepackage{amsfonts}%
\usepackage{amssymb}%
\usepackage{graphicx}
\usepackage[affil-it]{authblk}
\usepackage{hyperref}
\usepackage{textcomp}

\usepackage{lipsum}

\newcommand\blfootnote[1]{%
  \begingroup
  \renewcommand\thefootnote{}\footnote{#1}%
  \addtocounter{footnote}{-1}%
  \endgroup
}

\begin{document}

\title{\bfseries\LARGE Origin and suppression of parasitic signals in Kagom\'{e} lattice hollow core fiber used for SRS microscopy and endoscopy}


\author[1]{Alberto Lombardini}
\author[1,2,*]{Esben Ravn Andresen}
\author[2]{Alexandre Kudlinski}
\author[3]{Ingo Rimke}
\author[1,*]{Herv\'{e} Rigneault}

\affil[1]{Aix Marseille Universit\'{e}, CNRS, Centrale Marseille, Institut Fresnel, UMR 7249, 13397 Marseille Cedex 20, France}
\affil[2]{Universit\'{e} Lille, CNRS UMR 8523 - Laboratoire de Physique des Lasers Atomes et Molécules, F-59000 Lille, France}
\affil[3]{APE Angewandte Physik \& Elektronik GmbH, Haus N, Plauener Str. 163-165, D-13053 Berlin, Germany}
\affil[*]{Corresponding authors: esben.andresen@ircica.univ-lille1.fr, herve.rigneault@fresnel.fr}
\date{April 28, 2017}
\twocolumn[
  \begin{@twocolumnfalse}
    \maketitle
\begin{abstract}
Hollow core fibers are considered as promising candidates to deliver intense temporally overlapping picosecond pulses in
applications such as stimulated Raman scattering (SRS) microscopy and endoscopy because of their inherent low nonlinearity compared to solid-core silica fibers. Here we demonstrate that, contrary to prior assumptions, parasitic signals are generated in Kagomé lattice hollow core fibers. We identify the origin of the parasitic signals as an interplay between the Kerr nonlinearity of air and frequency-dependent fiber losses. Importantly, we identify the special cases of experimental parameters that are free from parasitic signals, making hollow core fibers ideal candidates for noise-free SRS microscopy and endoscopy.
\end{abstract}
  \end{@twocolumnfalse}
]\blfootnote{\textcopyright 2017 Optical Society of America. One print or electronic copy may be made for personal use only. Systematic reproduction and distribution, duplication of any material in this paper for a fee or for commercial purposes, or modifications of the content of this paper are prohibited.}

Stimulated Raman scattering (SRS)~\cite{Bloembergen_67} has received much attention since its implementation in microscopy~\cite{Freudiger_08,Nandakumar_09} as it provides a label free contrast mechanism with chemical sensitivity that is identical to spontaneous Raman scattering. SRS imaging has been implemented at video rate~\cite{Saar_10} and successfully applied to a variety of topics in cell biology and biomedical sciences~\cite{Cheng_15} to appear today as a viable platform with strong potential in biology and medicine. Although SRS \textit{per se} does not suffer from intrinsic non-resonant background such as the one found in coherent anti-Stokes Raman scattering (CARS)~\cite{Zumbusch_99}, the implementation of SRS is generally accompanied by parasitic signals stemming from other quasi-instantaneous nonlinear optical processes such as two-photon absorption~\cite{Fu_07}, cross phase modulation (XPM)~\cite{Samieni_12} and thermal effects~\cite{Lu_10}. When implemented with excitation beams propagating in free space, it has been previously demonstrated that SRS microscopy can be made free of these spurious signals using stimulated Raman gain and opposite loss detection ~\cite{Berto_14}. 
Implementations that involve propagation of the excitation pulses in optical fibers (i.e. the fiber delivery for SRS microscopes and SRS endoscopes) are much more prone to generate parasitic signals. Numerous attempts to perform CARS fiber beam delivery~\cite{Wang_06,legare_06,Jun_10,Wang_11} or probe/endoscopic imaging~\cite{Balu_10,Deladurantaye_14} have been reported. SRS fiber delivery~\cite{Brustlein_11} or scanning probe~\cite{Saar_11} have been less studied. But in general CARS and SRS using optical fibers have been hampered by the strong background signals arising from the wave mixing interactions in the fiber silica core. We concentrate here on Kagom\'{e} lattice hollow core fibers~\cite{Couny_06} (KL-HCFs) that are more suitable than band gap hollow core fiber~\cite{Knight_98}. Indeed, the broad spectral transmission band of Kagom\'{e} fibers has been exploited recently for development of Kagom\'{e}-based Raman probes~\cite{Ghenuche_12}, demonstrating that hollow core fibers lend themselves well to spectroscopic applications utilizing a single excitation beam, like spontaneous Raman. 

CARS and SRS are a different matter; in~\cite{Brustlein_11} we reported CARS and SRS using hollow core fibers that are known to exhibit extremely weak nonlinear effects because light propagates in the air core~\cite{Humbert_04}; although Raman spectral features from the sample could be retrieved, these signals sat on a plateau whose origin was uncertain.

In this Letter, our aim is to understand the origin of the residual parasitic signals in hollow core fibers, a topic that has important ramifications for the implementation of SRS beam delivery and endoscopy. Most importantly, we identify the experimental parameters leading to the complete cancellation of the residual parametric signals.

Fig.~\ref{Fig1:Fiber}(a) shows a SEM micrograph of the silica KL-HCF that we designed and fabricated;
light confinement inside the Kagom\'{e} core is not the result of photonic band gap, but is due to a mechanism of inhibited coupling between the cladding modes and the guided core modes~\cite{Couny_07}. 
The fiber has an outer diameter of 278~$\mu$m and a core diameter of 20~$\mu$m [Fig.~\ref{Fig1:Fiber}(b)] that confines the light in a single mode [Fig.~\ref{Fig1:Fiber}(c)]. The transmission window extends from 730~nm to 1100~nm with loss close to 5~dB/m (Fig.~\ref{Fig1:Fiber}(d)). The relatively high overall attenuation is mainly due to an unoptimized core contour shape \cite{wang2011} and small core size, for such KL-HCF. The attenuation curve also exhibits a strong wavelength dependance within the transmission band, which is due to anti-crossings with large cladding resonances \cite{travers2011}. This is a typical feature of KL-HCF having a non-optimized core surround \cite{travers2011,benabid2006}. The group velocity dispersion (GVD) is of the order of few ps/nm/km over the full transmission window. This low GVD assures that there is no pulse broadening over the 1~m fiber length propagation for the picosecond pulses used here.      

\begin{figure}[htbp]
\centering
\fbox{\includegraphics[width=.95\linewidth]{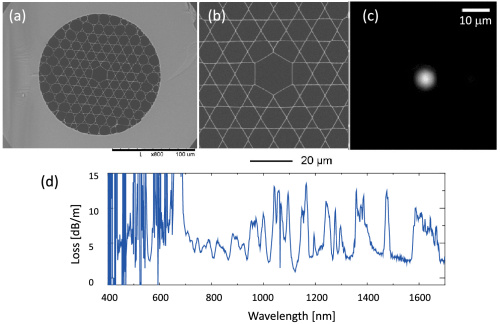}}
\caption{Kagom\'{e} lattice fiber: SEM micrographs of the full fiber (a) and close look at the fiber core (b). Mode intensity at 800~nm (c). Fiber loss spectra (d).}
\label{Fig1:Fiber}
\end{figure}

\begin{figure}[htbp]
\centering
\fbox{\includegraphics[width=.95\linewidth]{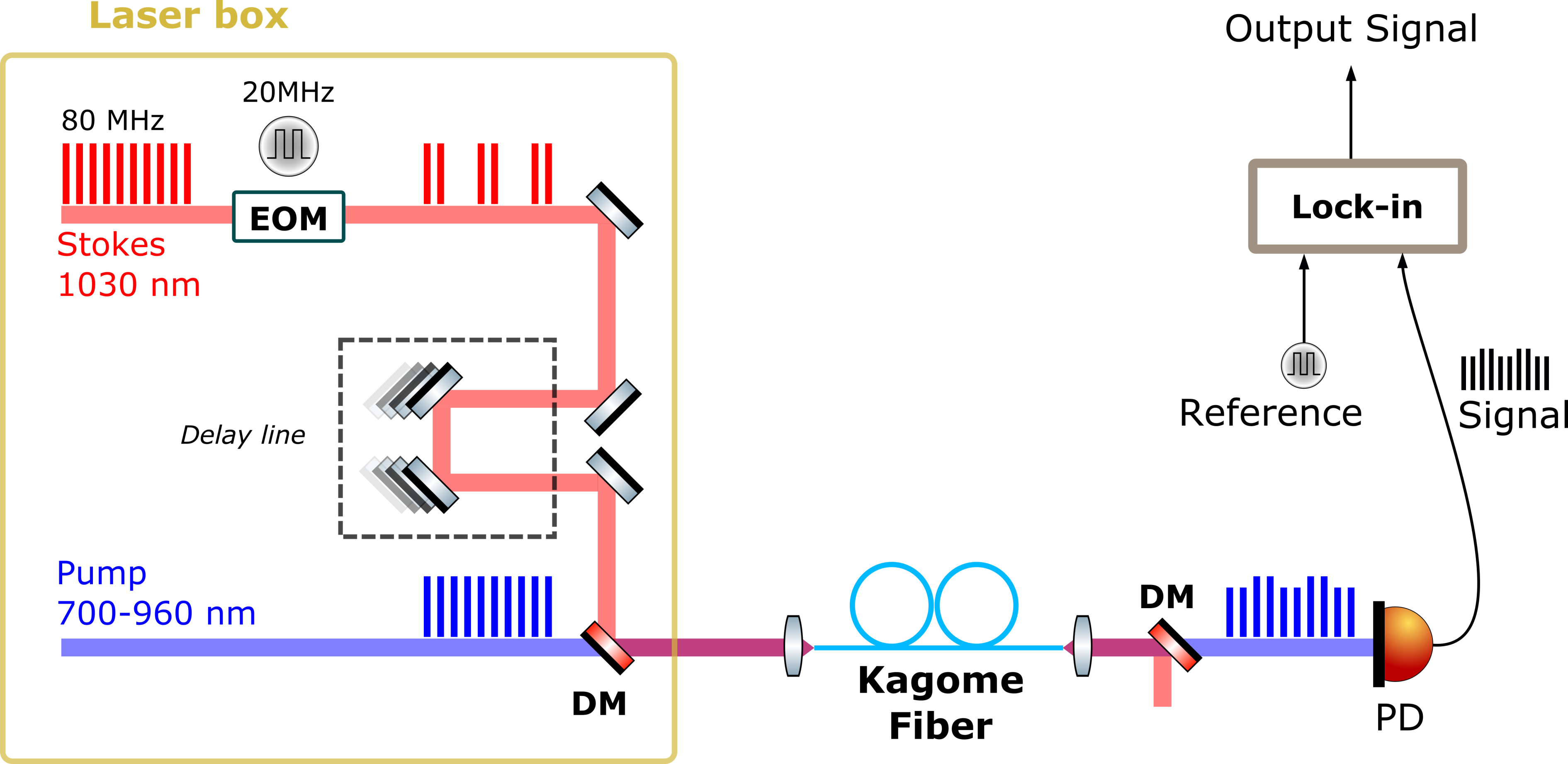}}
\caption{Setup: EOM electro-optical modulator, DM dichroic mirror, PD photodiode. The 20~MHz Lock-in reference signal is derived from the 80~MHz laser repetition with a frequency divider included into the picoEmerald{\texttrademark}~S laser system.}
\label{Fig2:Setup}
\end{figure}

Fig.~\ref{Fig2:Setup} is a schematic of the setup we have built to study the parasitic signals in KL-HCF. 
The setup is basically a setup for measuring the modulation transfer from the Stokes beam to the pump beam. 
As such, it mimicks the SRS excitation scheme known as the stimulated Raman loss (SRL) scheme.
In the SRL scheme pump and Stokes pulses interact to probe the sample Raman gain that is imprinted as a modulation transfer from a modulated beam (the Stokes) to a non modulated beam (the pump)~\cite{Freudiger_08}. 
The Stokes pulse train at 1030~nm is amplitude modulated with and electro-optical modulator (EOM) at 20~MHz and the high frequency pump beam (tunable from 700~nm to 960~nm) is unmodulated. The two beams, temporally and spatially overlaid, are provided by a picoEmerald{\texttrademark}~S from Angewandte Physik \& Elektronik GmbH (A.P.E). This laser system delivers 2~ps pulse duration and 80~MHz repetition rate suitable for SRS micro-spectroscopy. After propagation in 1~m of KL-HCF, the Stokes beam is filtered out using two laser line filters centered at 1030~nm, leaving only the pump beam on the detector. The pump beam is enlarged with a negative lens (not shown) to overfill the detector's sensitive area with a power of 40~mW. The detection at the EOM modulation frequency (20 MHz) of the pump beam is done by the A·P·E lock-in amplifier (LIA). The LIA integration time was set to 20~$\mu$s and its signal was set positive for modulation in phase with the modulated Stokes beam. It is possible to retrieve the fractional modulation in the pump beam (the amount of modulated optical power over the total pump power) from the LIA output signal $Signal_\mathrm{LIA}$ using the formula:
\begin{equation*}
\Delta I_P/I_P = \frac{Signal_\mathrm{LIA}}{K\cdot P_{\mathrm{det}}}
\label{eq fract mod lockin}
\end{equation*}
where $\mathrm{K=5\cdot 10^{5}~mV/mW}$ is a constant of the LIA, valid in conditions of LIA maximum gain (43.5~dB) and post amplification on, and $\mathrm{P_{det}}$ is the total pump power incident on the detector that should be kept lower than 50~mW to avoid saturation. We control the temporal overlap between pump and Stokes beams using at the KL-HCF output with a cross-correlator (pulseCheck USB{\texttrademark}, A.P.E). 

\begin{figure}[htbp]
\centering
\fbox{\includegraphics[width=.95\linewidth]{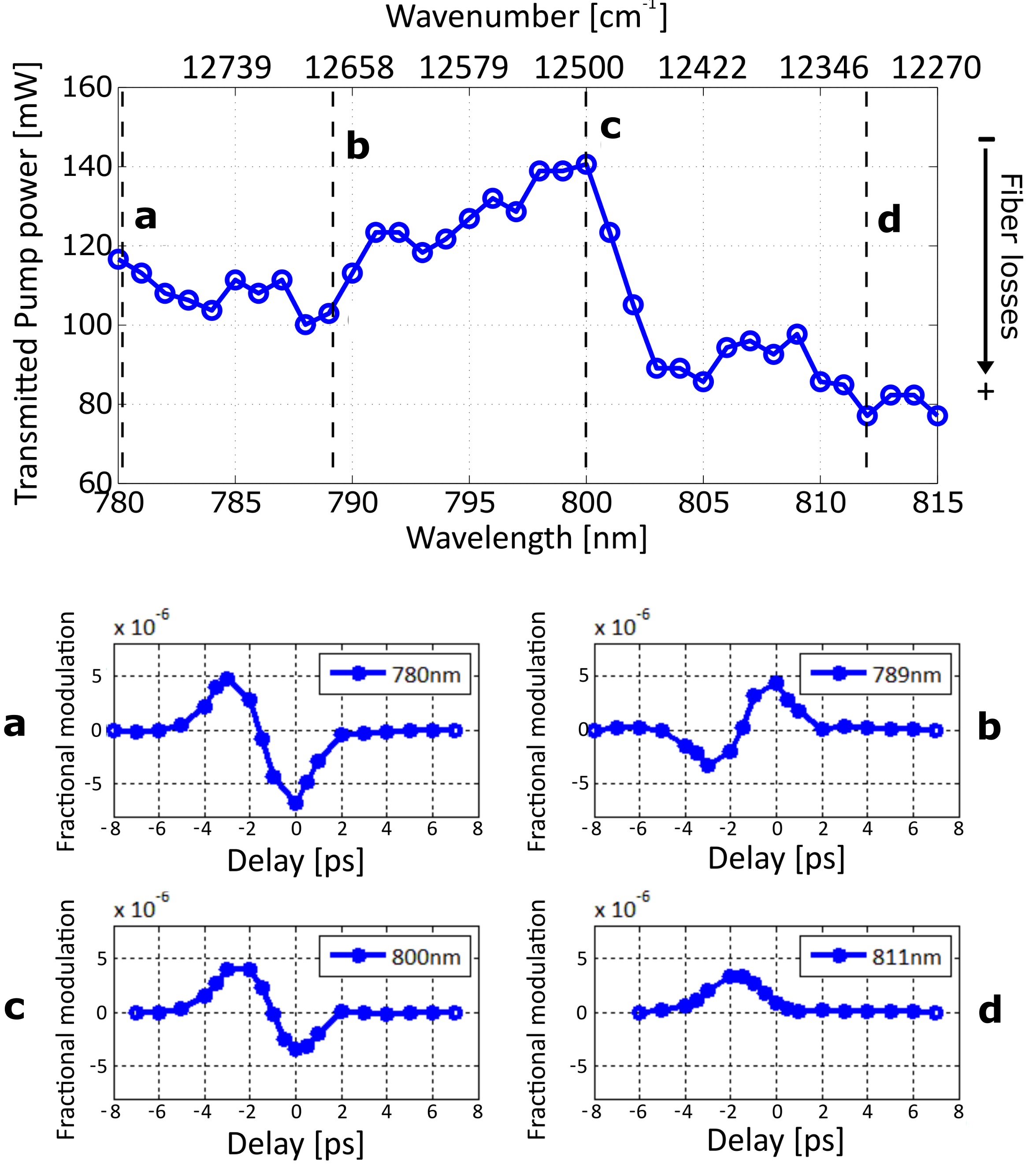}}
\caption{Transmission spectrum of the hollow core fiber and its fractional modulation with the delay between the pump and Stokes pulses at wavelength 780~nm (1), 789~nm (2), 800~nm (3) and 812~nm (4).}
\label{Fig3:Exp_results}
\end{figure}

Fig.~\ref{Fig3:Exp_results} shows the transmission of the KL-HCF as measured on the experimental setup when scanning the pump wavelength from 780~nm to 815~nm at constant input power of $\sim$~200~mW. 
The transmission is found to be highly wavelength dependent. Knowing the pointing stability of the laser pump beam (<~100~$\mu$rad~/~100~nm), we rule out any change in the coupling conditions and attribute this pump transmission power dependency to fiber losses only.
The measured parasitic signals [Fig.~\ref{Fig3:Exp_results}(a)-\ref{Fig3:Exp_results}(d)] as a function of delay at wavelengths (a), (b), (c) and (d) are of the order of few $10^{-6}$ (convention: The Stokes pulse is leading at negative delays). Unexpectedly, their lineshapes resemble dispersive lineshapes that can be positive or negative depending on  wavelength and delay. We have confirmed this behavior in other KL-PCF with slightly different designs (data not shown).    

\begin{figure}[htbp]
\centering
\fbox{\includegraphics[width=.95\linewidth]{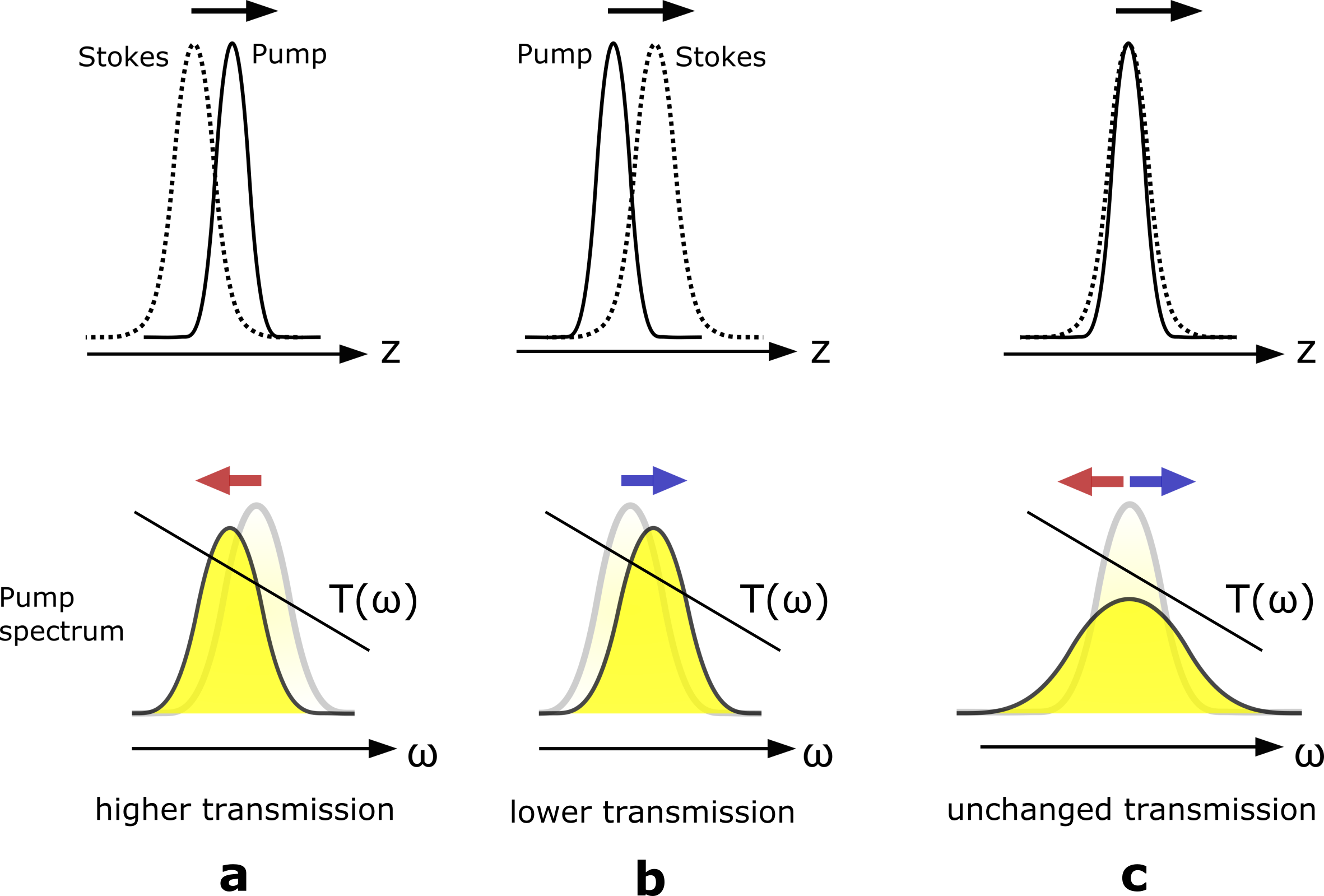}}
\caption{XPM at work between Stokes and pump pulses: Depending on the relative delay the pump pulse is redshifted (a), bueshifted (b) or broaden (c). Depending on the shape of the fiber transmission loss $T(\omega)$ (here we consider as an example with negative slope) this leads to higher (a), lower (b) or unchanged (c) pump transmission.}
\label{Fig4:Cartoon}
\end{figure}

To explain the magnitude and shape of the observed parasitic signals we propose a combined effect of cross phase modulation (XPM) and frequency dependent loss. Let us remind that the Kerr effect (refractive index change with the pulse intensity temporal profile $I(T)$) leads to a frequency shift $\delta\omega(T)$ that depends on the sign slope of the pulse intensity $\delta\omega(T)\propto-\frac{\partial I(T)}{\partial T}$. Such that in self phase modulation (SPM), the leading edge of the pulse ($\mathrm{\partial I(t)/\partial T>0}$) is shifted to lower frequencies ($\delta\omega<0$), while the trailing one is shifted to higher frequency. 
Fig.~\ref{Fig4:Cartoon} presents a intuitive cartoon, in the time and spectral domains, when XPM between the Stokes and the pump pulses is at work. Similar to SPM, XPM modifies the spectra of the pulses, but its effects changes with delay. When the Pump beam is leading [Fig.~\ref{Fig4:Cartoon}(a)] the slope of the Stokes causes the Pump spectrum to shift to lower frequencies (redshift), whereas when it is trailing Fig.~\ref{Fig4:Cartoon}(b) it shifts to higher frequencies (blueshift). When the two pulses overlap Fig.~\ref{Fig4:Cartoon}(c), similar to SPM, the pump spectral profile is broadened.
Now let us consider a frequency dependent loss medium such as the KL-HCF, depending if XPM is active or not (i.e if the Stokes pulse is present or not) the pump will experience different loss related to its time dependent frequency shift Fig.~\ref{Fig4:Cartoon} leading to a positive, negative or zero parasitic signal in SRS.

To further confirm this explanation, we have conducted numerical simulations to solve the nonlinear Schr{\"o}dinger equation (NLSE) using the split-step Fourier method and to compute the propagation of short pulses in optical fibers~\cite{Weideman_86, Sinkin_03}.
We describe the field as being composed by a Gaussian pulse (pump) and an hyperbolic secant pulse (Stokes), which corresponds to the experimental situation (data not shown). We shift the spectrum so that it is centered on the pump central frequency, which leads to the following form for the input field: 
\begin{equation}
\begin{split}
E_t=E_P^0\cdot \mathrm{e}^{(-2,77\cdot t^2/\tau_P^2)+i\cdot\alpha t^2}\\
 + E_S^0\cdot \mathrm{sech}(t/\tau_S)\cdot \mathrm{e}^{[i\cdot 2\pi\cdot(\nu_P-\nu_S)\cdot (t-t_{\mathrm{in}})]}
\end{split}
\end{equation}
where $\tau_P$=1800~fs is the FWHM of the Gaussian pulse, and $\tau_S=1.13$~ps. The initial time delay between pulses is set with the parameter $t_{\mathrm{in}}$. The frequencies $\nu_P$ and $\nu_S$ are the frequencies of the Pump and Stokes beam, corresponding to 800~nm and 1031~nm. The time axis is discretized into 3~fs increments and the frequency axis into  0.3~cm$^{-1}$ increments. The field propagates in 1~m of fiber, which is divided in 200 steps over which the effect of dispersion and nonlinearity are computed. The parameter $\alpha$ is an initial chirp term on the Gaussian pulse that is used to match the time-spectral duration with the pulse used for experiment. The Kagom\'{e} dispersion is negative and low, for which we impose a value of $\beta_2=-3000$~fs$^{2}$/m. We consider air as the nonlinear medium ($n_2=3\cdot 10^{-23}$~m$^2$/W) and the effective area of the propagating mode $A_{\mathrm{eff}}=220$~$\mu$m$^2$.
The frequency-dependent losses are applied along with the dispersion and the nonlinearity in each step of the propagation. After the final step the total pump output power $P_{\mathrm{out}}^{\mathrm{pump}}$ is found by integrating the output spectrum in a range around the pump central frequency. The modulation induced by XPM is then found as the difference between $P_{\mathrm{out}}^{\mathrm{pump}}$ in the SPM case (only pump pulse in the fiber) and the XPM case (with the Stokes pulse).

\begin{figure}[htbp]
\centering
\fbox{\includegraphics[width=.95\linewidth,height=.88 \textheight,keepaspectratio]{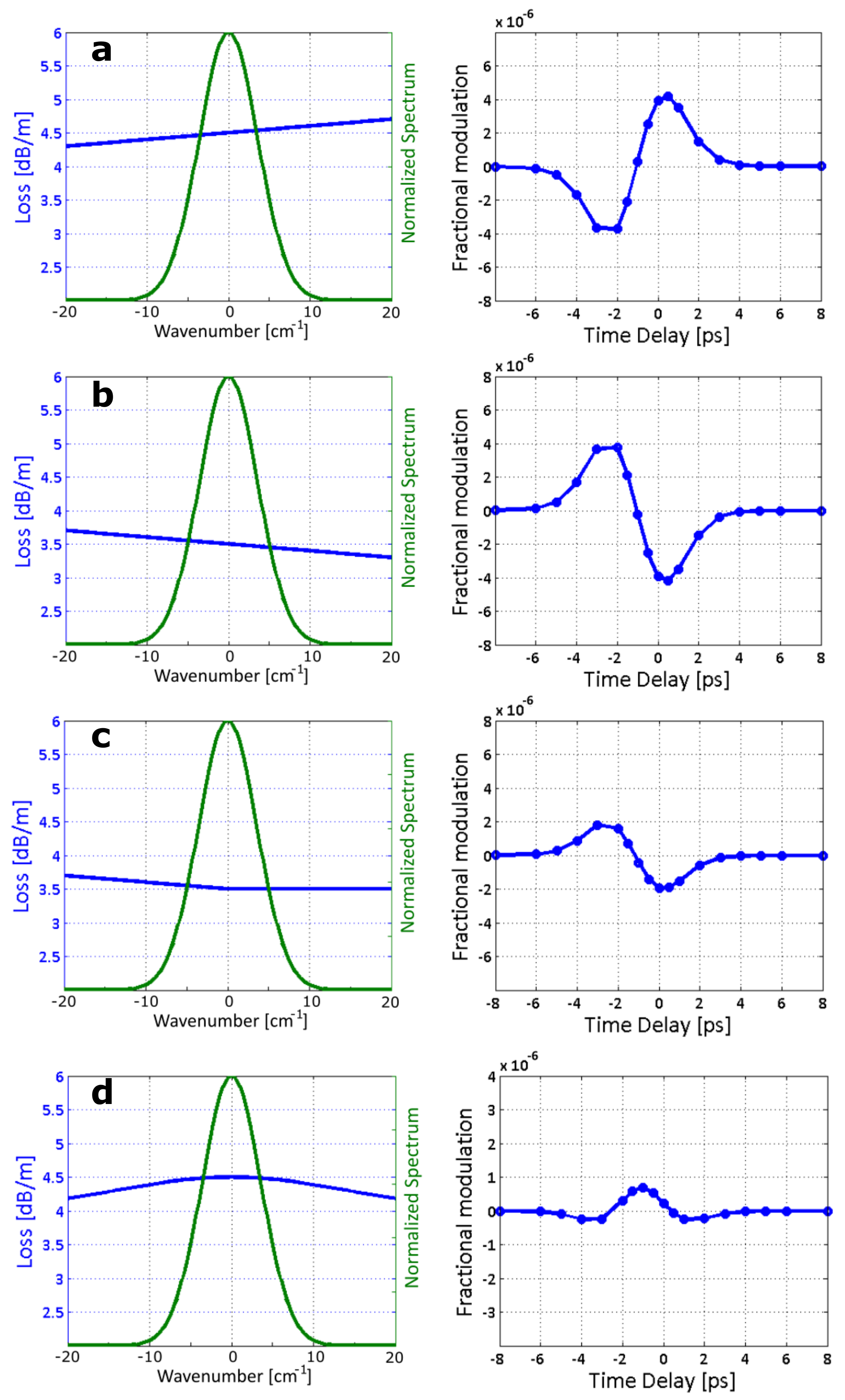}}
\caption{Simulation of various fiber spectral loss profiles (blue lines) over the pump pulse spectral bandwidth (green line) and their associated fractional modulation versus the time delay between the pump and Stokes pulses. The time delay convention is the same as in Fig.~\ref{Fig3:Exp_results}}
\label{Fig5:Sim_results}
\end{figure}

Figure~\ref{Fig5:Sim_results} shows the result of the simulation for different spectral loss profiles (blue lines) over the pump pulse spectral bandwidth (green line) and the associated parasitic signal versus delay. 
The time delay convention is the same as in Fig.~\ref{Fig3:Exp_results}. Fig.~\ref{Fig5:Sim_results}(a), (b), (c) and (d) correspond to situation similar to cases (a), (b), (c) and (d) in Fig.~\ref{Fig3:Exp_results}, respectively. The XPM asymmetric spectral broadening (see Fig.~\ref{Fig4:Cartoon}) taking into account the fiber losses leads to dispersive-like line shapes with the delay, in good agreement with measurements Fig.~\ref{Fig3:Exp_results}. These include the cases of a linear profile with slope $\approx 0.3$~dB/nm. Note that the parasitic signal vs delay curve is reversed when the slope goes from positive to negative. The parasitic signals are on the order of few $10^{-6}$ and have a local maximum and one minimum when the frequency shift is maximized. In between there exists a delay where it is zero. When the loss spectrum is flat on one side (case (c)), there is still a positive, then negative modulation as a function of the delay. When the pump beam is centered on a local maximum of the loss spectrum (case (d)),  the fractional modulation vs delay shows a global maximum, centered at the delay point giving maximum broadening, in qualitative agreement with Fig.~\ref{Fig3:Exp_results}(d). If the absorption maximum is due to a resonance with the glass structure, a significant part of the guided light will be interacting with glass whose nonlinear index is 1000 times that of air. In such a case, our simulation (which assumes a nonlinear index equal to that of air) would give only a lower estimate of the parasitic signal. 
In the case of a flat transmission spectrum the modulations are very small, on the order of $10^{-15}$ (data not shown).  
It is important to note that the above mentioned parasitic signals in KL-HCFs are still $10^{4}$ lower from what can be found in large mode area (LMA) fibers and that KL-HCF remain excellent candidates to deliver temporally overlapping ps pulses for SRS microscopy~\cite{Brustlein_11} and endoscopy. Finally, the minimal fractional modulation which is possible to detect with the A.P.E LIA module is around $5\cdot 10^{-7}$ for the conditions used in the experiment (50~mW on the LIA detector, 20~$\mu$s integration time) which corresponds to the shot noise limit. Using the picoEmerald{\texttrademark}~S, which is shot noise limited~\cite{Rimke_14}, the KL-HCF affects the SRS ultimate sensitivity by a factor of 10 only (for this particular integration time). From Fig.~\ref{Fig3:Exp_results} cases (a), (b) and (c), there exists a time delay around $\sim$-2~ps where the parasitic signal is identical to zero; future applications of beam delivery through KL-HCF should exploit this fact in order to retain shot noise limited sensitivity in SRS. 

In conclusion, we have shown that KL-HCF are subject to a residual parasitic signal (SRS background) that results from a combination of cross phase modulation between the pump and Stokes pulses in the fiber air core and frequency dependent fiber loss. This parasitic signal is of the order of few $10^{-6}$ and depends on the delay between the pump and Stokes pulses and the spectral loss profile in the vicinity of the non modulated beam wavelength. Used in an SRS scheme, this additional fractional modulation degrades by a factor of 10 only the ultimate sensitivity brought by shot noise limited laser systems and detection electronic. KL-HCF with lower loss will lead to a proportional reduction of the parasitic signals. Even so, we have also established that the parasitic signal can be completely canceled by correctly choosing experimental parameters, bringing hopes for shot noise limited SRS fiber endoscopy.

\bigskip
\textbf{Funding.} European Commission (EC) (607842 FINON ITN-2013); Agence Nationale de la Recherche (ANR) (10-INSB-04-01, 11-IDEX-0001-02, 11-INSB-0006, 14-CE17-0004-01, 11-EQPX-0017, 11-LABX-0007); Institut National
de la Santé et de la Recherche Médicale (Inserm) (PC201508);European Regional Development Fund (ERDF); Centre National de la Recherche Scientifique (CNRS) (IRCICA USR 3380);

\end{document}